\documentclass[prb,preprint,superscriptaddress,showpacs]{revtex4}
\usepackage{revsymb}
\usepackage[dvips]{graphicx}
\usepackage{color}
\usepackage{here}
\usepackage{amsfonts}
   
\begin{document}
   
\title{{\LARGE On study of nonlinear viscoelastic behavior of red 
blood cell membrane.}}
   
\author{ {\large {\bf Horacio Castellini}} }
\email{hcaste@ifir.edu.ar}
\affiliation{{\rm Dpto\@. de F\'{\i}sica, F\@.C\@.E\@.I\@.A\@.,
Pellegini 250, 2000 Rosario}}

\author{{\large {\bf Bibiana Riquelme}} }
\email{bibiana_riquelme@yahoo.com.ar}
\affiliation{{\rm Fac\@. Cs\@. Bioqu\'{\i}micas y Farmac\'euticas, 
Suipacha 531, 2000 Rosario}}

\author{{\large {\bf Patricia Foresto}} }
\affiliation{{\rm Fac\@. Cs\@. Bioqu\'{\i}micas y Farmac\'euticas, 
Suipacha 531, 2000 Rosario}}

\begin{abstract}
The linear viscoelastic behavior of the red blood cell membrane of 
mammal and human was studied in previous works proposing different 
experimental methods to determine their viscoelastic parameters. 
In the present work the nonlinear component of dynamic viscosity of 
the red blood cell membrane by nonlinear time series analysis is used. 
For such aim, it obtained time series of test {\em in vitro} of samples 
of humans and rats red blood cells using the Erythrodeformeter in 
oscillating regime. The signal filtrate suppresses any linear behavior as 
well as represented by a system of linear ordinary differential equations. 
The test shown as much in humans as in rats resonance frequencies 
associated to an attractor of unknown nature independently of 
excitation in the physiological range. 
The preliminary studies shown that attractor could be correspond to a 
complex form bull. These results allow to extend the present knowledge 
on dynamic of the cellular membrane to similar stimulus which happens 
in the blood circulation and it will allows to make better 
models of the same one.
  
\end{abstract}

\maketitle

\section{Introduction}
In previous works\cite{riq} we have demonstrated that the 
diffractometric method 
is adapted to make the complex viscoelastic analysis of human red blood 
cell (RBC) and turns out to be a very useful tool to investigate the 
properties of these cells. The complex viscoelasctic parameters of human 
RBC determined by laser diffractometric have been and continue 
being very useful for the detection of alterations produced in 
different levels from the RBC membrane or by diseases such as the 
diabetes, some anemia or the arterial hypertension, thus also like 
the induced ones {\em in vitro} by synthetic lectins, 
glicosilation or polications\cite{riq1} \cite{riq2}.
   
The measurements made with the Erythrodeformeter in oscillating regime 
have allowed to verify that for frequencies and shear rate within 
the physiological rank the RBC deformation varies sinusoidal since it 
makes the tension of cut with a shift phase $\phi$ that is function of 
the oscillation frequency, it is to say that the RBC for that rank of 
shear stress can actually be like a linear viscoelastic materials. 
Nevertheless the results show the existence of underlying 
nonlinear viscoelastic component.

The method of the recurrent plots (RP) allows to visualize recurrent 
patterns in a time series of data. This technique was in first time 
propose by Eckman {\em et al\@.}\cite{ekm} in 1987 and to find complex and 
hidden correlations between the data. At the moment it is applied in 
the diverse fields as much of physics\cite{zbi}, chemistry\cite{rus}, 
the economy\cite{hol} etc.
   
\subsection{Theoretical Foundations} 
An RP is an injective application of a single reconstructed trajectory 
to Boolean matrix where each value pairs $y_i$ and $y_j$ coming from
a time series is related with a numerical pair $(i,j)$ called recurrence 
point. Let us consider {\em N} join values from a time series given by:
\begin{equation}
Y=\{ y_0, y_1, \ldots, y_{N-1} \}
\end{equation}
with {\em N} large enough in order to evaluate the embedding dimension
$d > 2$ by using the false nearest neighbor algorithm and the time 
delay $\tau >1$ by looking at the relative minimum in the mutual 
information\cite{fra}. 
Following Takens' embedding theorem the dynamics can be appropriately
mapped by the phase space trajectory reconstructed by using the time 
delay vectors:
\begin{equation}
A_k=\{ y_k, y_{k+\tau}, y_{k+2 \tau}, \ldots, y_{k+ (d-1) \tau} \} 
\end{equation}
with {\em false-neighbours} algorithm\cite{ken} it is possible to obtain 
dimension embedding $d$.

Its symmetric recurrence matrix is:
\begin{equation}
R_{(i,j)}=\Theta(\delta_h-||A_i-A_j||_\infty) -
\Theta(\delta_l-||A_i-A_j||_\infty)
\end{equation}
where $\Theta$ is the {\em Heaviside} function. This means a RP is 
built by comparing all delay vectors with each other. When
$\delta_l < ||A_i-A_j||_\infty < \delta_h$ the $R_(i,j)$ component is 
set to ``{\bf 1}'' value, otherwise it is set to ``{\bf 0}'' value. The 
interval $[\delta_l,\delta_h]$ is known as {\em threshold corridor}. 
The choice of this interval is critical, if too large produce a saturation of
RP including irrelevant points, and if too narrow loses information.
A good criterion is the suggested by Zbilut\cite{zbi} in which the 
scale fractal zone is used to choose the possible values of this passage. 
This criterion is used in this work to find the advisable values 
of $\delta_l$ y $\delta_h$. Webber {\em at al\@.}\cite{zbi1} in order to 
characterize and analyze RP introduced a set of quantifiers which 
are collectively called {\em recurrence quantified analysis} (RQA). 

{\bf a)} The first one is the  
{\em percentage of recurrence} (\%REC), it is defined as:
\begin{equation} \texttt{\%REC}=100 \, \frac{N_r} {N_t} \end{equation}
where $N_t=\textrm{dim}(R)$ (every possible points) and
$N_r$ is number of recursive points given by
\begin{equation}
N_r= 2 \, \# \{ (i,j) / R_{(i,j)} > 0  \, \textrm{and} \, i<j \}.
\end{equation}
where $\#$ is set cardinal function.
The slope of linear region in the S-shaped \%REC vs corridor width
is the correlation dimension. 

{\bf b)} The second quantifier is called 
{\em percentage of determinism} (\%DET), it is defined as:
\begin{equation} \texttt{\%DET}=100 \, \frac{N_l}{N_r}  \end{equation}
where $N_l$ is called the number of periodic dots given by
\begin{equation}
N_l = 2 \, \# \{ (i,j) /  (i,j) \in d_c(k,b), \, i<j, 
\forall \, c, \, k, \, b>0. \}
\end{equation}
and a periodic line with length $b$, origin $k$ and zone $c>0$ is
defined as:
\begin{equation}
d_c(k,b) = \{ (i,i+c) /  \prod_{i=k}^{k+b} R_{(i,i+c)} >0. \}
\end{equation}
The \%DET is related with the organization of RP.
   
{\bf c)} The third quantifier is called
{\em entropy} (S), it is defined as the first rate cumulants of periodic lines:
\begin{equation} 
S=-\sum^H_{b=1} P_b \log_2(P_b) 
\end{equation}
The label entropy is just that, a label, not to be confused
with Shannon's entropy.
$H$ is the length of maximum periodic line and 
$P_n > 0$ is the relative frequency of periodic lines with $b>0$. 
Webber assume that S is related with Shannon's entropy if and only
if the system is chaotic and the embedding dimension large enough.

{\bf d)} The fourth and last quantifier is $H$. 
Eckman {\em et al\@.}\cite{ekm} claim that line lengths on RP are directly 
related to inverse of the largest positive Lyapunov exponent. 
Short lines values are therefore indicative of chaotic or stochastic behavior.
   
Recently Castellini and Romanelli\cite{cas} have developed an 
effective algorithm to evaluate RQA applied to the study of 
transitions in chaotic chemical reactions. 
The objective of the present work is to extend that 
methodology to the study of the viscoelasticity nonlinear of the RBC 
membrane as much for human RBC beings like of RBC rats. 
The election of this animal must to that the linear answer of the RBC is 
similar one of RBC human and it is used habitually for studies of 
diseases induced in alive. In this work it is tried to determine 
if there are significant differences between both species when nonlinear 
behavior by means of the RQA as by power spectra studies as much within 
the physiological rank.

\section{Method and Material}
\subsection{Human RBC}
The human RBC was obtained from whole of healthy givers and anticoagulated 
blood with EDTA. The samples were maintained to $4^\textrm{o}$C until 
their use which was made according to the recommendations of the 
{\em Society International of Clinical Hemorheology} and within 
the 48 later hours to its extraction.
\subsection{Rat RBC}
The originating blood of 10 male rats of line "m" was extracted post 
morten of the anticoagulated heart and with heparin. Previously 6 of the 
rats were injected by intraperitoneal route with aloxane which induces 
diabetes and the other four were considered like controls. 
Glicemia of the rats dealt with aloxane went of 2.90$\pm$0.80 to the 
48 hours of the injection and it stayed within that rank during the 
two weeks of the work.
\subsection{Preparation of the suspension}
The whole blood sample was suspended in isotonic viscous means constituted 
by a solution to 5\% w/v of polivinil-pirrolidone 
(360 PVP Sigma, p.m. 360 kDa) in buffer saline phosphate 
(PBS: 0.150 M NaCl, and 0.005 M; pH 7.4$\pm$0.05; 295$\pm$8 mOsmol/kg). 
The viscosity of means was fit to 22.0$\pm$0.5 cp to $25^\textrm{o}$C.
\subsection{System of measurement}
A thin layer of RBC suspension is placed between both concentric horizontal 
discs and of the Erythrodeformeter. The superior disc is fixed and the 
inferior one is movable. An adjustable power supply provides to the motor 
of a stationary voltage or an oscillating sinusoidal voltage. 
In stationary regime the inferior disc spins at constant speed. 
In dynamic conditions the inferior disc spins at speeds that can 
vary sinusoidal way with six pre-established frequencies: 
0.5, 1.0, 1.5, 2.0, 2.5, and 3 Hertz. 
A laser beam perpendicularly crosses the sample to RBC suspension 
producing a diffraction pattern that is to circulate for RBC without 
shear stress and elliptical under shear stress.
\subsection{Time series analysis}
The obtained temporary series of data was filtered to eliminate all 
well-known sinusoidal frequency response. 
For it a frequency groove was used\cite{van}. 
The resulting remainder $r(t)$ again it was filtered with 
a {\em Savitzky-Golay}\cite{fla} filter 
of sixth order which allows to eliminate white noise phenomena in the data. 
It is possible to emphasize that this it is a F\@.I\@.R\@. 
filter (Finite Impulse Response) and not a I\@.I\@.R\@. filter 
(Infinite Impulse Response) reason that attractor behavior is not affected 
by this type filtering\cite{bro}. 
Then the data to null average $\mu=0$ and variance 
unity $\sigma=1$ were standardized. 
The filtered temporary series contains the behavior nonlinear 
looked for with an optimal relation signal noise.

\begin{center}
\begin{tabular}{c}
Table I \\
\begin{tabular}{|c|c|c|c|c|}
\hline
Excitation frequency (Hz) & $f_1$ (Hz) & $f_2$ (Hz) & $f_3$ (Hz) & $f_4$ 
(Hz) \\
\hline
0,5 & 0,921 & {\bf 1,043} & 1,087 & 1,14 \\
1,0 & 1,234 & 1,322 & 1,82 & {\bf 2,087} \\
1,5 & 1,108 & 1,171 & {\bf 1,234} & 2,861 \\
\hline
\end{tabular}
\end{tabular}
\end{center}

\begin{figure}[ht]
\begin{center}
\includegraphics[width=12cm, height=15cm]{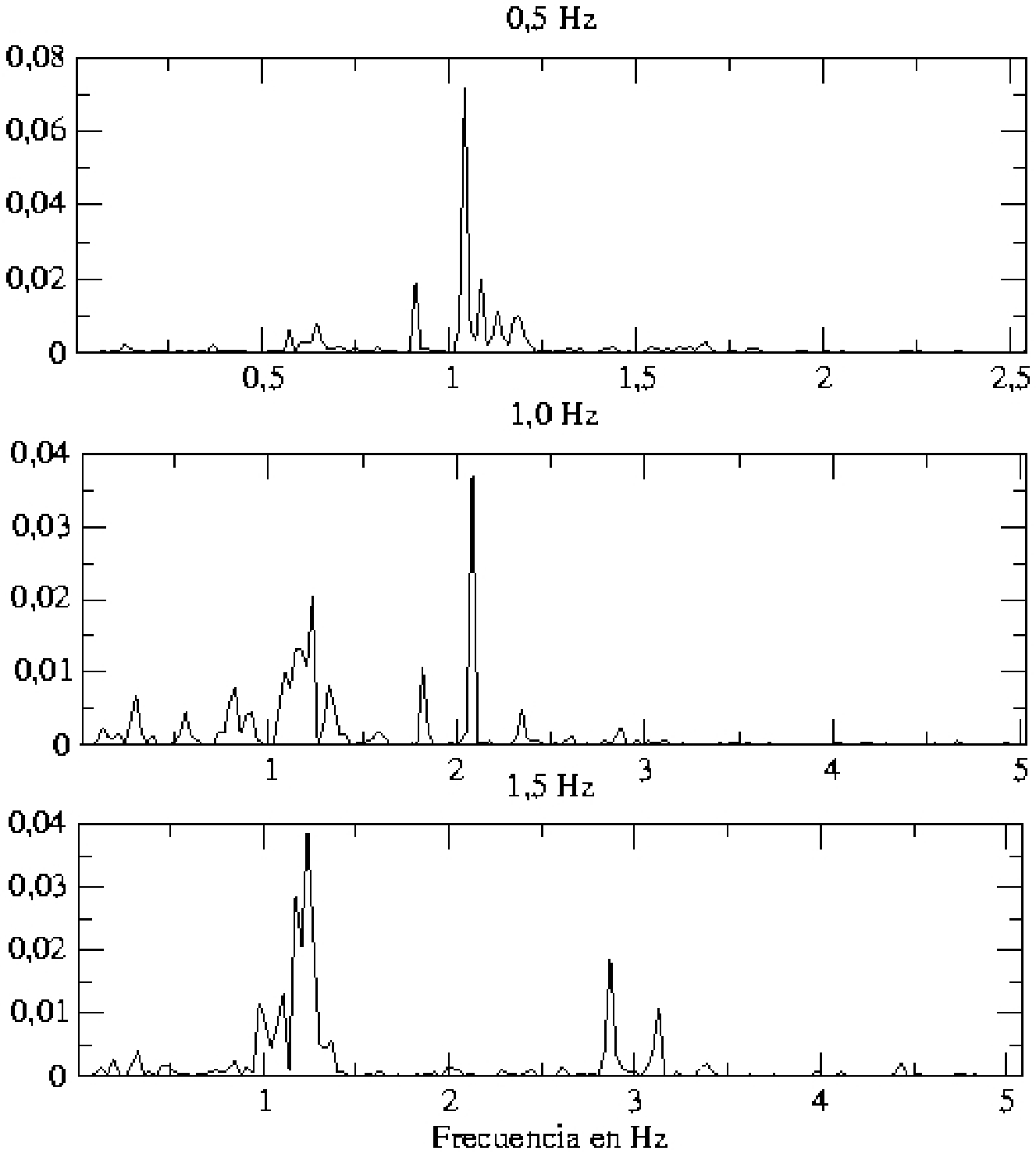}
\end{center}
\caption{Power spectra of Human RBC in different excitation frequencies. 
It is posbible to appreciate the different sharp that correspond to 
their frequencies of resonances. Where their values depend on the 
excitation frequency.}
\label{fig:1}
\end{figure}

As it can be seen in figure-\ref{fig:1} for different physiological 
frequencies of excitation they appear a few of resonance sharp. 
In table I is the numerical value of the frequency for these sharp in 
human RBC. We only write the values of more dominant frequencies. 
The values emphasized in bold correspond to greater amplitude sharp. 
With different values from frequency were analogs results in red RBC 
of healthy and aloxane rats occur. 
This suggests the presence of a attractor that is sensible to small 
changes of its environment and the species to which investigates. 
The RQA allows to quantify the attractor topologic behavior.
   
Before applying algorithm RQA it was determined as the minimum of the 
mutual information is the average delay $\tau$ investigating. 
The value obtained in sampling passage turned out to be $\tau=10$. 
The concrete value in seconds depends on the sampling frequency and 
it is of little relevance in this type of analysis. 
Previous to calculate embedding dimension by {\em false-neighbors} 
method the Theiler's window in sampling passage was considered whose 
value was $\tau_t=15$. The embedding dimension calculated was $d=5$.
Based on these data the time delay vector series was generated to apply RQA.

\section{Results and discussion}
\begin{figure}[ht]
\begin{center}
\includegraphics[width=12cm, height=15cm]{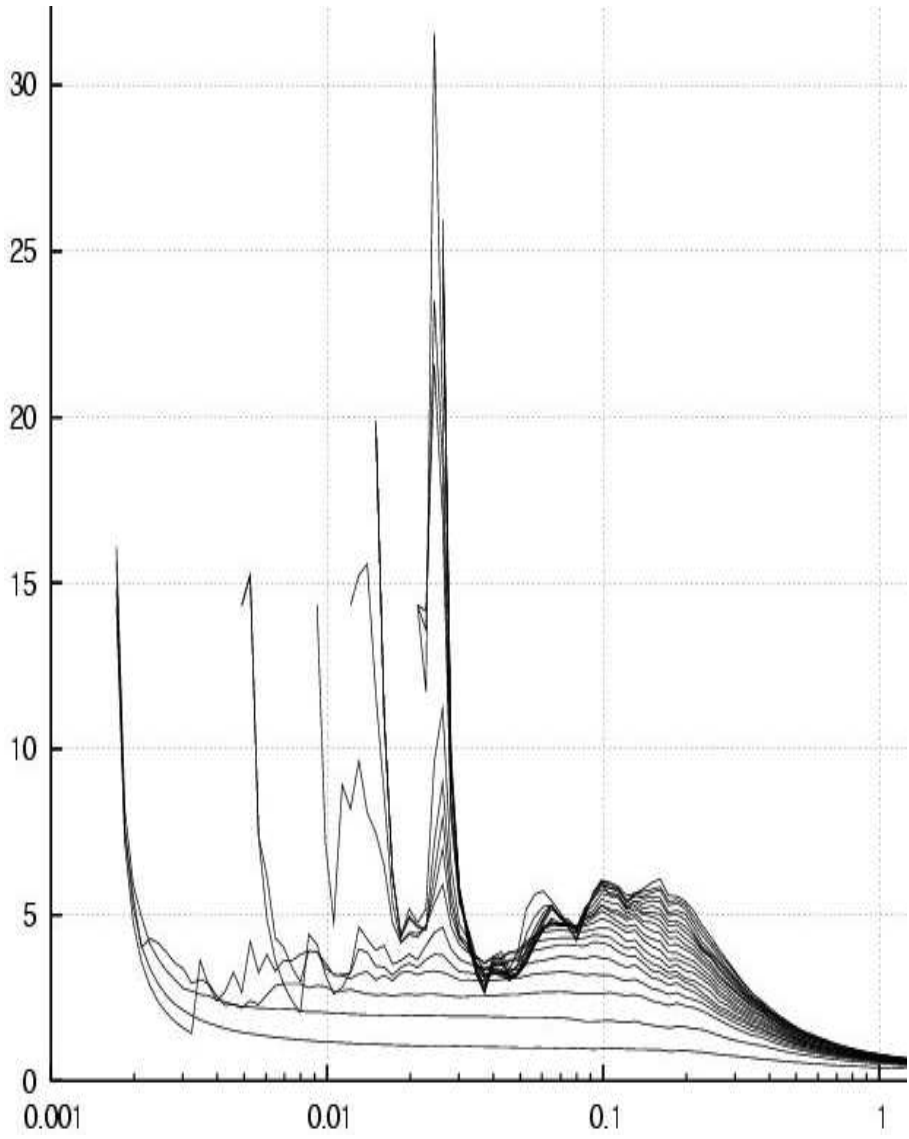}
\end{center}
\caption{The correlation dimension $C_2$ by means of the Grassberger and 
Procaccia's algorithm. It is posbible to appreciate the region in 
which $C_2$ estimation becomes independent of $d$ and $\epsilon$ 
is reduced}
\label{fig:2}
\end{figure}

\begin{figure}[ht]
\begin{center}
\includegraphics[width=12cm, height=15cm]{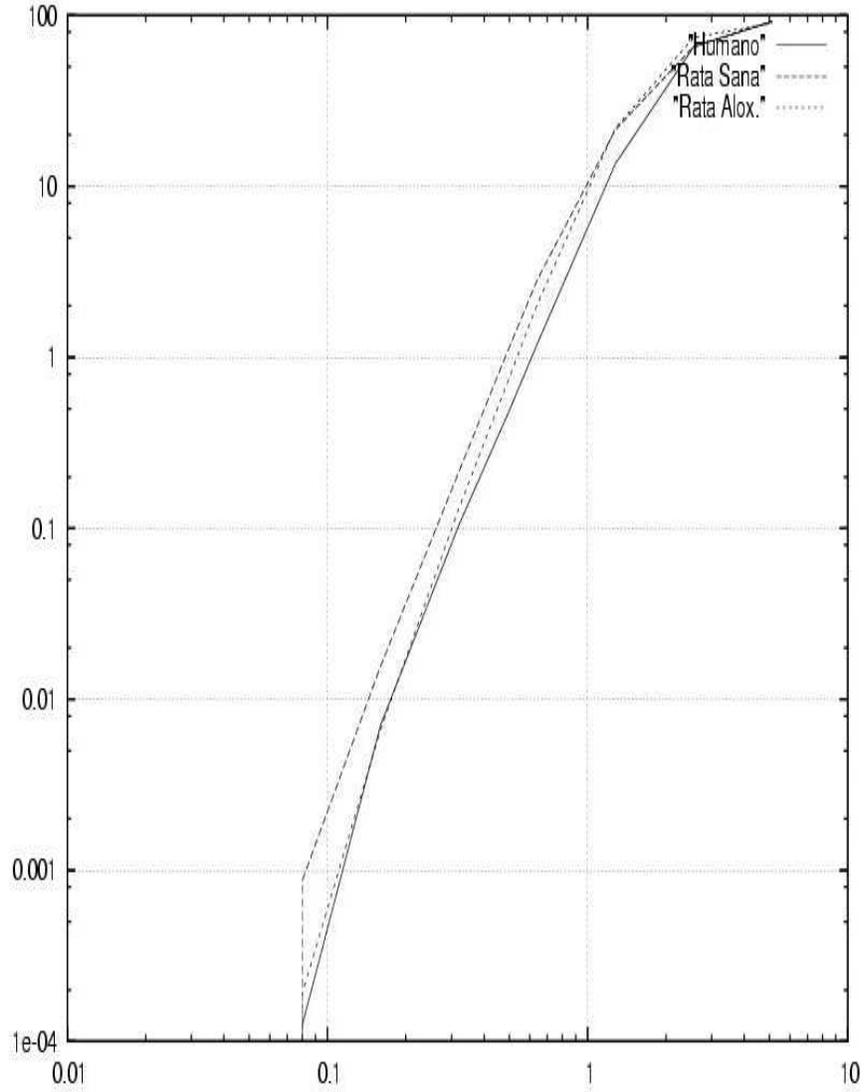}
\end{center}
\caption{The correlation dimension $C_2$ by means of RQA analysis. The slope 
of this curve in S-shaped is $C_2$. 
It is possible appreciate that curves corresponding to 
healthy individuals are almost parallel in S-shaped zone. }
\label{fig:3}
\end{figure}

The correlation dimension $C_2$ by means of the Grassberger and 
Procaccia's\cite{gra} algorithm was calculated. 
The tests made with human RBC gave to a value of $C_2=3.48 \pm 0.68$.
As the region in which can be appreciated in figure-\ref{fig:2} the 
estimation of $C_2$ becomes independent of $d$ and $\epsilon$ is reduced 
giving a not convincing result.
However when RQA was used, figure-\ref{fig:3}, to evaluate the dimension 
of correlation and the wide of passage the S-shaped zone is independent 
not only from the frequency of excitation also from the species in 
individual in the case of RBC of healthy individuals. 
This suggests that topology of the attractor is unique with a calculated 
dimension of $C_2=3.5 \pm 0.3$ that agrees in probability with the value 
before obtained. An interesting detail to consider is that in the case 
of aloxanized rats this dimension is increased to a value 
$C_2=4.0 \pm 0.3$. Based on this result the wide one of the calculated passage 
was $\delta_h-\delta_l=0.64$ . 
Then RQA was used to calculate the maximum length H and entropy S as 
it is shown in table II.

\begin{center}   
\begin{tabular}{c}
Table II \\
\begin{tabular}{|l|c|c|}
\hline
Sample & $H$ (Dots) & $S$ (bits/ut) \\
\hline
Human to 0,5 Hz & 285 & 1,82 \\
Human to 1,0 Hz & 241 & 1,66 \\
Human to 1,5 Hz & 165 & 1,66 \\
Healthy Rat to 0,5 Hz & 343 & 2,21 \\
Healthy Rat to 1,0 Hz & 845 & 2,51 \\
Healthy Rat to 1,5 Hz & 843 & 2,13 \\
Rat Dealt with to 0,5 Hz & 304 & 1,77 \\
Rat Dealt with to 1,0 Hz & 234 & 1,30 \\
Rat dealt with to 1,5 Hz & 319 & 1,62 \\
\hline
\end{tabular}
\end{tabular}
\end{center}

\section{Conclusions}
Results obtained previously in samples of rats dealt with aloxane, 
they did not show to alterations in the linear hemorheologic parameters 
like the deformability and the aggregation due to the diabetes induced by 
this treatment\cite{cam}. 
Nevertheless, in this work we found that the dimension of 
correlation $C_2$ measured with RQA allowed to observe differences in 
relation to the controls. On the other hand as it is appraised in 
Table II the entropy shows differences between healthy human 
RBC and healthy rats RBC indicating that these last ones display a 
non-linearity as ratifies parameter H. However it is interesting to 
emphasize the sensitivity of the method when the rats are affected by 
diabetes induced by the aloxane because the entropy is much smaller 
for these RBC of altered rats that the controls. 
Equal result is observed in H. The reasons for this behavior until have 
not been now deciphered considering these restlessness for future studies.
In this study one concludes that the parameters determined by means 
of the use of the RQA are great sensitivity to detect alterations in 
RBC membrane no detectable by other techniques. 
Consequently these results would be of great applicability in the 
biomedical investigation.


\end{document}